# Dimensionality-dependent electronic and vibrational dynamics in low-dimensional organic-inorganic tin halides

*Yanmei He[1], Xinyi Cai[2], Rafael B. Araujo[3], Yibo Wang[1,4], Sankaran Ramesh[1], Junsheng Chen[5], Muyi Zhang[2], Tomas Edvinsson[3], Feng Gao[2, *], Tönu Pullerits[1, *]*

[1] Division of Chemical Physics and NanoLund, Lund University, P.O. Box 124, 22100 Lund, Sweden

[2] Department of Physics, Chemistry, and Biology (IFM), Linköping University, SE-581 83 Linköping, Sweden

[3] Department of Materials Science and Engineering – Solid State Physics, Uppsala University, Box 534, Uppsala SE-75121, Sweden

[4] Department of Nuclear Science and Technology, Nanjing University of Aeronautics and Astronautics, Nanjing 211106, China

[5] Nano-Science Center & Department of Chemistry, University of Copenhagen, Universitetsparken 5, 2100 Copenhagen, Denmark

*Corresponding author: tonu.pullerits@chemphys.lu.se; feng.gao@liu.se

**Abstract**

Photo-induced dynamics of electronic processes in materials are driven by the coupling between electronic and nuclear degrees of freedom. Here we construct 1D and 2D organic-inorganic tin halides to investigate the functional role of dimensionality to exciton-phonon coupling (EPC) and exciton self-trapping. The results show that the 1D system has strong EPC leading to excitation-independent self-trapped exciton (STE) emission, while the 2D system exhibits over ten times weaker EPC resulting in free exciton emission. By performing femtosecond transient absorption experiments, we directly resolve the room-temperature vibrational wavepackets in the 1D system, some of which propagate along the STE potential energy surface. A combination of wagging and asymmetric stretching motions (~106 cm$^{-1}$) in tin iodide is identified as such a mode inducing exciton self-trapping. While no room-temperature wavepackets are observed in the 2D system. These findings uncover the interplay between the dimensionality-dependent EPC and electronic/nuclear dynamics, offering constructive guidance to develop multifunctional organic-inorganic metal halides.

**Introduction**

Solution-processed low-dimensional hybrid organic-inorganic metal halides have attracted significant interest due to their promise in enabling efficient optoelectronic devices.[1, 2, 3, 4, 5] The soft ionic crystal structure of these materials gives rise to complex interactions between the electronic structure and lattice vibrations, named exciton-phonon coupling (EPC). This tunes the optical and electronic properties of the material and provides pathways for exciton/carrier relaxation and dissipation.[6, 7, 8, 9] For instance, a profound exciton-phonon interaction in two-dimensional (2D) layered lead halide perovskites (LHPs) usually leads to the formation of large polarons, where the polarization of the local environment can shield hot carriers thereby slowing down the thermalization of the non-equilibrium photoexcited carrier population via electron/hole-longitudinal optical (LO) phonon scattering or Auger heating process.[10, 11, 12] Alternatively, since in ionic systems like LHPs, the Fröhlich coupling of the electrons and phonons engages mostly the low-momentum LO phonons, the hot carrier cooling selectively excites only those phonons, thereby rapidly establishing thermal equilibrium between the selected LO modes and the charge carriers. This can drastically slow down the further cooling of the hot carriers—an effect called hot phonon bottleneck.[13, 14] The peculiar EPC thus inspires the development of versatile LHPs for possible future efficient hot-carrier solar cells.[15] In one-dimensional (1D) and zero-dimensional (0D) organic-inorganic metal halides, EPC often results in self-trapped exciton (STE) emission with a large Stokes shift compared to the 2D counterparts. Usually, the STE emission has high quantum efficiency promoting their potential for developing white light-emitting diodes (WLEDs).[16, 17, 18, 19]

The femtosecond pump-probe techniques, including impulsive stimulated Raman scattering (ISRS),[20] impulsive vibrational spectroscopy,[21] ultrafast electron diffraction,[22] and femtosecond transient absorption (fs-TA) spectroscopy,[23] allow for the real-time observation of the photoexcited coherent phonon dynamics since a short laser pulse can excite a coherent phonon wavepacket. By using these methods, the carrier/exciton-phonon interactions in the 2D LHPs have been comprehensively explored. For example, the studies of coherent phonon dynamics in LHPs, including (PEA)$_2$PbI$_4$ and (FPT)PbI$_4$, reveal that the low-frequency vibrational modes (< 60 cm$^{-1}$) involving the bending and stretching vibration in lead halide octahedra, play a crucial role in polaron formation and in that way modify the hot carrier cooling process.[11, 21, 23, 24] However, the details of how dimensionality influences EPC and thereby controls the optoelectronic properties of organic-inorganic metal halides are still elusive. Especially, the possible functional role of EPC in low-dimensional organic-inorganic tin halides is yet to be understood. Tuning the EPC strength by modifying the dimensionality of these materials might not only manipulate the electronic transitions but also affect the coherent phonon dynamics on a microscopic scale. Therefore, understanding the nature of dimensionality-dependent EPC in organic-inorganic tin halides is important for tailoring their optical and electrical properties for solar cells, LEDs, and even lasers.

In this work, we tune the ligand concentration to prepare the 2D and 1D Dion-Jacobson (DJ)-type organic-inorganic tin halides dressed by soft ionic lattice, leading to distinctly different coupling strengths between the lattice vibrations and electronic structure in these two systems. As we will show, the dimensionality has consequences for the properties of excitons and their dynamics – in 2D system free excitons (FEs) dominate while in 1D structure STEs are formed. Using fs-TA spectroscopy under resonant excitation conditions, we reveal intrinsic differences between the FEs and STEs by comparing their excitation-dependent electronic and temperature-dependent vibrational dynamics. The STE population dynamics are independent of excitation intensity up to very high exciton densities ($> 2.9 \times 10^{20}$ excitation/cm$^3$/pulse), while in most 2D perovskites, Auger recombination appears above $1.0 \times 10^{18}$ excitation/cm$^3$/pulse. Moreover, coherent oscillations are observed in the 1D system at room temperature, while the coherence is absent in the 2D system. Based on Fourier transform analysis and theoretical modeling of oscillatory wavepacket dynamics, we establish a microscopic picture of vibrational dynamics in 2D and 1D systems. We conclude that the ligand-modified EPC significantly influences the phonon anharmonicity and lattice distortions in the tin iodide, altering also the electronic dynamics. These studies unravel the underlying mechanism for the dimensionality-dependent emission and shed light on the lattice dynamics in low-dimensional organic-inorganic tin halides.

**Results**

**Structural and photophysical properties**

DJ-type organic-inorganic tin halides ODASnI$_4$ and ODASn$_2$I$_6$ (ODA=1,8-octane-diamine) thin films were prepared by adjusting the molar ratios between the organic cations and the tin halides (Supplementary Note 1).[25, 26] The morphology, elemental composition and photoluminescence quantum yields (PLQYs) of these materials are shown in Supplementary Fig. S1-3, demonstrating their good quality. The powder X-ray diffraction pattern of ODASnI$_4$ thin film presents periodic peaks at 13°, 20°, 27° and 34° which are similar to that in 2D ODAPbI$_4$ while the periodicity is absent in ODASn$_2$I$_6$ thin film accompanied by the appearance of a new peak at 8° (Supplementary Fig. S4).[27] Fig. 1 displays the narrow emission of ODASnI$_4$ thin film and broad emission of ODASn$_2$I$_6$ thin film. Given the distinct emission features and diffraction patterns, we thus conclude that the 2D phase is dominant in ODASnI$_4$ thin film, while the ODASn$_2$I$_6$ thin film has a 1D nature. The real structures are probably disordered to some extent but the main feature, either being 1D or 2D, dominates these two systems. In the following text, we refer to them as 1D and 2D systems. In the absorption spectrum, we observed the excitonic peaks at 502/588 nm for the 2D system and 359 nm for the 1D system. The PL spectrum of the 2D system shows a relatively small Stokes shift of ~ 27 nm ($\lambda_{em}$ = 616 nm) with a weak low-energy tail. However, the 1D system has a significantly larger Stokes shift of ~251 nm ($\lambda_{em}$ = 610 nm). We also noticed a substantially longer PL lifetime and higher activation energy in the 1D system compared to the 2D one (Supplementary Fig. S5-6). These distinct photophysical properties suggest that the electronic transitions in 2D ODASnI$_4$ and 1D ODASn$_2$I$_6$ are dominated by FE and STE states, respectively.

Given the importance of the coupling between electronic and nuclear degrees of freedom for optical properties and dynamics of the system, we strive for a quantitative description of the exciton-phonon interaction strength. For that we use the Huang-Rhys ($S$) factors of the vibrational modes that we determine both via experiments and computations. First we analyze the temperature-dependent fluorescence line width as $FWHM(T) = 2.36\hbar\omega\sqrt{S \cdot (2n_T + 1)}$ where FWHM is the full width at half maximum, $n_T$ is Bose-Einstein occupation number, and $\hbar\omega$ is the effective mode energy (Fig. 1f).[28, 29] The obtained S factor for the 1D system is ~ 36 with the effective mode frequency of 160 cm$^{-1}$. We also analyzed the bandgap changes as a function of time via a combined molecular dynamics – electronic structure calculation. From the Fourier transform of the bandgap fluctuation autocorrelation function, we obtained a spectral density function that provides mode frequencies modulating the bandgap and the relative strengths of the corresponding S factors, as shown in Fig. 1g-i. The final $S$ factors were derived from an analogous expression as above where instead of fluorescence FWHM, the root mean square of the bandgap fluctuations was used (for more details see Supplementary Note 2).[30, 31, 32] The fluorescence linewidth analyses lead to a larger reorganization energy $\lambda = \sum_i S_i \hbar\omega_i$ than the calculated bandgap fluctuations because the higher energy tail of the spectral density function stretches beyond what is represented by the 3 modes that are used in the analyses. The general agreement of the two methods provides evidence for a strong displacement of the excited state nuclear equilibrium position compared to the ground state. This leads to the formation of the STE state and distorted lattice at the excited state of the 1D system causing the broad red-shifted emission. In the 2D system, the weak emission tail at 660 nm, coming from the surface trap states (SUT), makes the line-width analysis for the *S* factor unreliable (Supplementary Note 2 and Fig. S7). Below we will quantify the EPC strength of the 2D ODASnI$_4$ using the method derived from the oscillatory components in the fs-TA measurements.

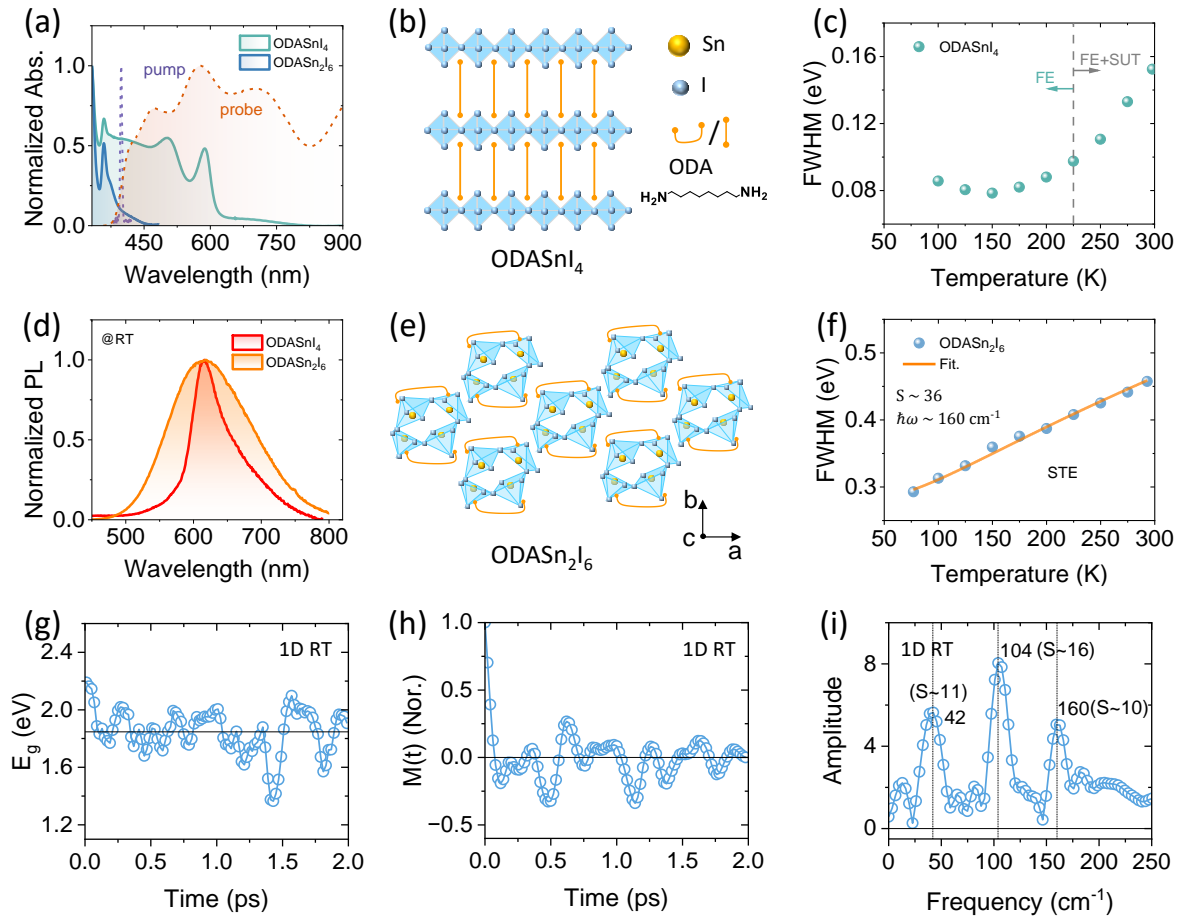

**Fig. 1 Structure, steady-state spectra, and exciton-phonon coupling estimation.** 2D ODASnI$_4$ and 1D ODASn$_2$I$_6$ thin films: (a) UV-vis absorption spectra. Pump, and probe spectra in fs-TA measurements; (b) and (e) Schematic crystal structure (In the 1D structure, the four corner-sharing [SnI$_5$]$^{3-}$ moieties align along the c-axis like a glide plane separated by ligands);[26, 27, 33] (c) and (f) Temperature dependence of FWHM; (d) PL spectra; For the 1D system: (g) Calculated energy gap time-correlation function; (h) Autocorrelation function M(t) of the energy gap; (i) Fourier transform spectrum of M(t). The Huang-Rhys factor and effective mode frequency are shown for comparison.

**Electronic dynamics**

TA measurements used excitation at 400 nm (~ 3.1 eV) and a broad white light probe allowing to collect detailed information on exciton dynamics. As shown in Fig. 2, the 1D and 2D systems show very different TA signals. In the 2D system, we identify two ground-state bleaching (GSB) bands at 506 and 584 nm, which are consistent with the excitonic peaks observed in the steady-state absorption spectrum in Fig. 1a. Three excited state absorption (ESA) bands accompany the GSBs. At the first 1 ps, the GSB bands show a slight red-shift (~ 3 nm), reflecting hot carrier cooling (Supplementary Fig. S8a-b). With increasing excitation density, the main band-edge bleach shows an obvious blue-shift because of the Burstein-Moss effect (Supplementary Fig. S8c).[34, 35] The previous studies demonstrate that the bandgap renormalization due to the photogenerated charge carriers, a so-called excitation-induced shift, can also affect the bandgap of the material, typically resulting in a red-shift of GSB peaks.[36, 37] These two effects partially compensate each other resulting in the observed slight GSB red-shift. In Fig. 2d, the TA spectrum of the 1D system shows a broad ESA signal across the probe wavelength region. Such a spectrum is consistent with the STE state seen in earlier studies.[38, 39]

By using the singular value decomposition (SVD) global analysis (GLA) method, we now quantify the excited state dynamics in these materials. In the 2D system, we obtained four components: $\tau_1$ ~ 210 fs, $\tau_2$ ~ 56 ps, $\tau_3$ ~ 620 ps, and $\tau_4$ > 5 ns (see Supplementary Fig. S9c). The ESA signal at > 625 nm rapidly decays with a few hundred fs lifetime, suggesting that the ultrafast component $\tau_1$ corresponds to the hot carrier cooling. We carried out the intensity-dependent TA measurements, shown in Fig. 2, Supplementary Fig. S9-11 and Table S2. The lifetime of the component $\tau_1$ becomes longer at high excitation intensities due to the hot phonon bottleneck.[3] Conversely, the time constants $\tau_2$ and $\tau_3$ decrease dramatically. We explain this behavior in terms of general second and third order decay. Indeed, at the lowest excitation intensity, the band-edge GSB kinetics show a linearity of $\Delta A^{-1}$ as a function of delay time at first 600 ps resulting from the second order process, while at higher intensities they show the linearity of $\Delta A^{-2}$ vs delay time (indicative for the third order process), see Supplementary Fig. S12.[34, 35] Given this behavior, the longer decay components should not be directly interpreted as simple well-defined rate processes but rather reflect a combination of the second-order Saha-Langmuir type carrier recombination (evidenced by $\Delta A^{-1}$ dependence), the third-order Auger recombination at higher excitation intensities (evidenced by $\Delta A^{-2}$ dependence), and carrier-trapping by the surface states (evidenced by the low-energy emission tail in the steady-state emission spectrum). In the 1D system, the GLA results also give four components, but with notably different time constants (Supplementary Fig. S13c). The precise value of the fastest component of 480 fs has considerable uncertainty since the first half-phase of the pronounced oscillatory signal on the same timescale disturbs it. Still, it is clearly present. We follow the model of two STE states in our previous work with refinements based on the new data and closer analyses.[25] For the fastest component, we need to consider the nature of the initial state directly after light absorption. The tightly packed 1D structures likely support coherent delocalization over at least a few "wires", then because of the strong EPC we expect that the initially delocalized state rapidly looses coherence and localizes analogously to the formation of excitonic polarons in molecular complexes.[40] We therefore assign the fastest component to the excited state localization together with the cooling and the initial phase of the STE formation. The assignment is further supported by the similarity in spectral shape between the two fastest components and the observed spectral shift to higher energies. This is expected from an ESA signal reflecting the initial state relaxing to lower energies, thereby increasing the ESA transition energy. The second component of 2.8 ps converts the initial broad single-band spectrum to a rather different form with two bands (see Supplementary Fig. S13c), indicating the formation of a new state – $STE_2$. The two longest components have similar spectra suggesting no further change in state character. The longest component of $\gg$ 5 ns has half the amplitude of the 47 ps component. This, together with the PL quantum yield of 37 % makes us suggest that the 47 ps component corresponds to a nonradiative decay channel due to some structural peculiarities that affect about 60% of the excitations, while the longest component is mostly radiative decay with $\gg$ 5 ns time constant. The kinetics of the 1D system do not show almost any intensity dependence – this result is also very different compared to the 2D system. It means that the localized STE state is so well shielded by the deformation of the local environment that they do not interact with each other even at the excitation density over $2.9 \times 10^{20}$ excitation/cm$^3$/pulse. Furthermore, the PL intensity increases linearly with the excitation fluences (Supplementary Fig. S19), demonstrating that no additional nonradiative recombination channel appears in the 1D system at higher excitation intensities. It also means that the PL in this system is mostly the first order process rather than the second order electron-hole pair recombination as in the bulk perovskite microcrystals.[41]

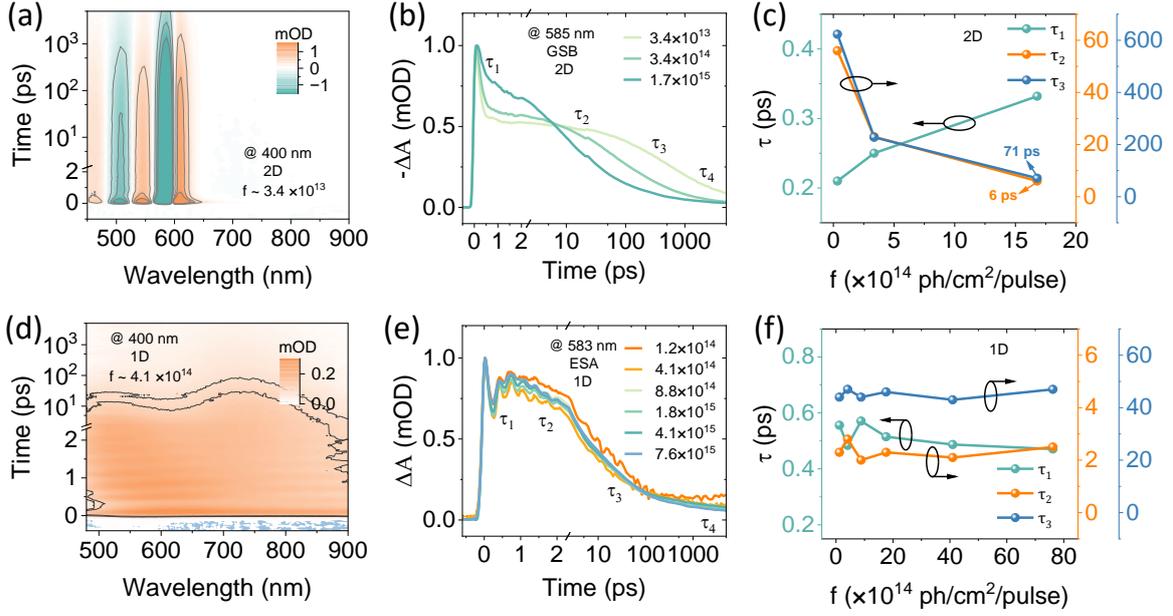

**Fig. 2 Exciton dynamics at different excitation fluences.** Pseudocolor representation of fs-TA spectra, temporal kinetics, and time constants extracted from SVD analysis of (a-c) 2D ODASnI$_4$ and (d-f) 1D ODASn$_2$I$_6$ thin films.

**Vibrational dynamics**

The ultrashort laser pulse can excite coherent vibrational wavepackets of the modes coupled to the electronic transition. Their generation mechanism can be described by a damped harmonic oscillator driven by the external force (details see Supplementary Note 4). The wavepackets oscillate on the potential energy surface with certain mode frequencies and the amplitude of these oscillations strongly depends on the exciton-phonon coupling strength in the materials. As shown in Fig. 2, a clear oscillatory signal appears in the 1D system at room temperature. Interestingly, no oscillations can be observed in the 2D system. Considering that the EPC and phonon dephasing are subject to temperature, we conducted fs-TA measurements also at 77 K and 200 K. The time-wavelength spectrograms of the differential absorption are shown in Fig. 3 and Supplementary Fig. S21-22. We found that at 77 K oscillations appear even in the 2D system. After subtracting the population dynamics from TA signals, we derive beating maps of oscillatory components (Supplementary Fig. S23). The vibrational modes are analyzed from the oscillatory signals in terms of a sum of damped cosine functions

$$\Delta A_{osc} = \sum_{i=0}^{n} A_i e^{-t/\tau_i} cos(2\pi v_i t + \varphi_i),$$

where $A$, $\tau$, $v$ and $\varphi$ are the amplitude, dephasing time, mode frequency, and phase of the i$^{th}$ oscillating component, respectively. The fitting results, presented in Supplementary Fig. S21-22, 24 and Table S3-4, reveal that the vibrational modes that appear in the TA dynamics are very different in the two systems. The dephasing time of the dominant phonon mode in the 2D system is over 1.4 ps, which is approximately 2.5 times longer than in the 1D system (~ 0.53 ps). By Fourier transforming the oscillatory signals, we uncovered the frequencies of vibrational modes and the amplitude of oscillations across the probe wavelengths, presented in Fig. 3c-d, Supplementary Fig. S25-26 and Table S5. The fitted vibrational spectrum allows to assign five modes in the 1D system: M2 (25 cm$^{-1}$), M3 (43 cm$^{-1}$), M4 (78 cm$^{-1}$), M5 (103 cm$^{-1}$) and M6 (123 cm$^{-1}$). At higher temperature a low-frequency mode M1 (15 cm$^{-1}$) appears and a slight upwards shift of the mode frequencies takes place accompanied by the broadening of the bands due to the enhanced phonon-phonon scattering. Especially, the M5 frequency changes to 106 cm$^{-1}$ at room temperature. In comparison, the 2D system shows only three vibrational modes at 77K, all below 50 cm$^{-1}$, with the most intense peak at 47 cm$^{-1}$ (Fig. 3f). The V1 (3 cm$^{-1}$) represents the very low-frequency phonon wing and is probably partially overlapping with the DC component. No such low-frequency component is observed in the 1D system. All modes in the 2D system are strongly damped and not visible at room-temperature dynamics. We propose that the significant differences in vibrational modes of 1D and 2D systems can be due to the variations in ligand concentration, resulting in a distinct degree of lattice distortion and phonon anharmonicity.[20]

We mentioned that the EPC strength estimation of 2D ODASnI$_4$ from the temperature-dependent PL spectra is not reliable because of the emission of a surface trap state. To facilitate the comparison of EPC strength in these systems, we estimate the Huang-Rhys factor *S* for the 2D system by adopting a displaced harmonic oscillator model that has been used for quantum dots and perovskites to obtain the EPC strength from the coherent wavepacket TA signal.[11, 20, 42, 43, 44] In this analyses, the EPC strength is proportional to the amplitude of the oscillatory signal ($\Delta A_{osc}$) which is related to the lattice reorganization energy ($\lambda$) through $\Delta A_{osc} = \lambda \cdot \frac{dOD}{dE}$ (for details see Supplementary Note 4). We focus on the vibrational modes at the band-edge GSB signal (595 nm). Using $\frac{dOD}{dE}$ from the normalized absorption spectrum, we obtain the *S* factor for all the modes and the results are summarized in Supplementary Table S6. The calculated *S* factors of the V3 and V4 modes are 0.8 and 1.3, respectively. This method cannot be used for the 1D system since we do not observe the GSB bands in the detection window. Given the previous results, we noted that the *S* factor of the 2D system is drastically smaller than that of the 1D system (~ 36), clearly demonstrating that the dimensionality strongly influences the exciton-phonon interaction in these materials.

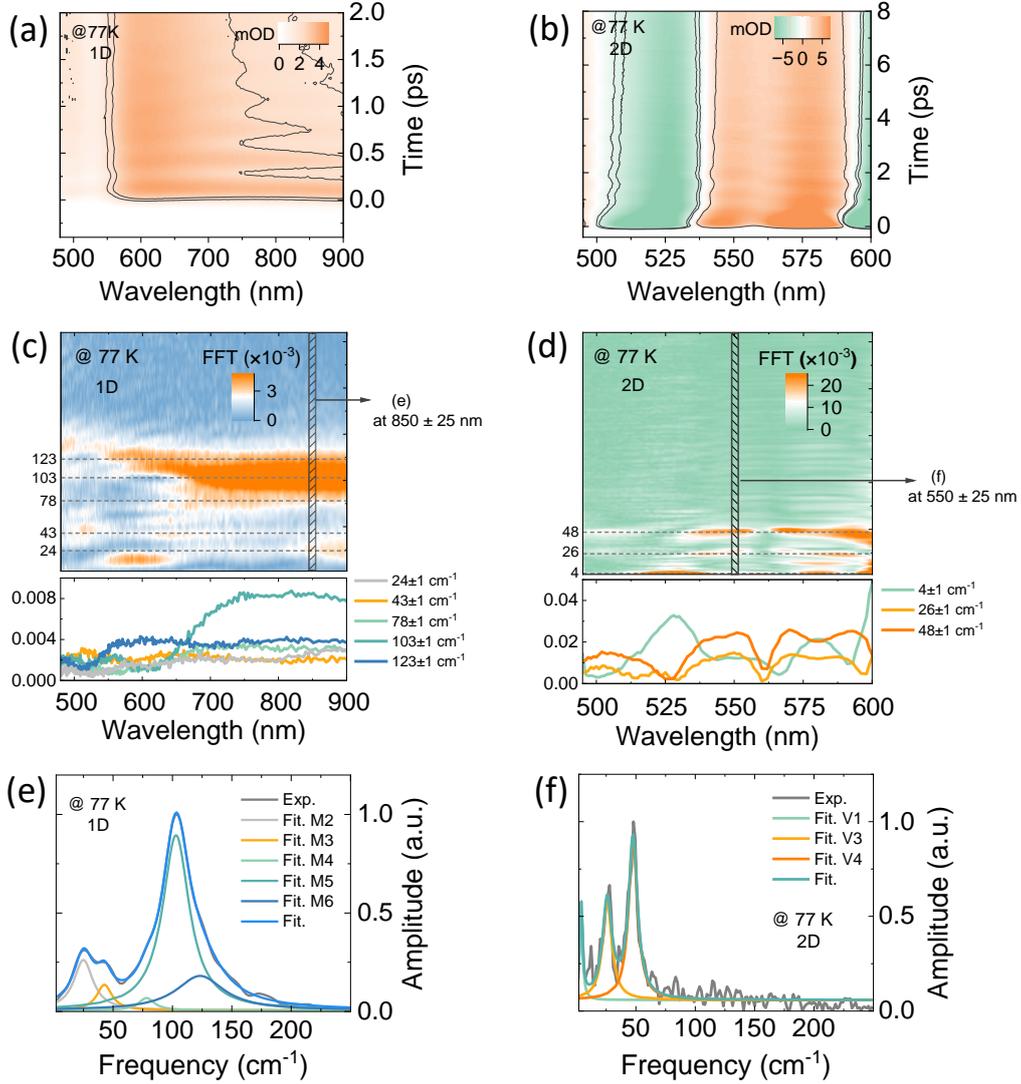

**Fig. 3 Vibrational mode analysis of oscillatory wavepacket at 77 K.** Pseudocolor representation fs-TA spectra of (a) 1D and (b) 2D systems. Probe wavelength resolved vibrational frequencies of (c) 1D and (d) 2D systems directly obtained by Fourier transforming the differential TA spectrum. Vibrational spectra of (e) 1D (probed at 850 ± 25 nm) and (f) 2D systems (probed at 550 ± 25 nm) with the Lorentzian fitting results. V and M are used to distinguish the modes in these two systems.

To properly discern the nature of the strong EPC in the 1D system and assign the observed coherent vibrational modes, we conduct the first-principles calculations of ODASn$_2$I$_6$ using density functional theory (DFT) with finite differences approach and density functional perturbation theory (DFPT) within the Vienna ab initio Simulation Package (VASP).[45]

Additionally, the temperature-dependent velocity autocorrelation function (VACF) is simulated by the Ab initio molecular dynamics (AIMD) method. As shown in Supplementary Fig. S27, the vibrational density of states (VDOS) spectrum has prominent peaks below 250 cm$^{-1}$, with the strongest peak in the range of 80-110 cm$^{-1}$. The FFT analysis of the electronic bandgap autocorrelation function in Fig. 1 also reveals a dominant vibrational mode at 104 cm$^{-1}$, which is consistent with the M5 mode obtained from the TA dynamics. The electronic transition mainly involves the tin and iodide atoms, suggesting these modes are closely related to the atomic motions in tin iodide units (Supplementary Fig. S28). The temperature-dependent analysis of VACF further shows a shorter dephasing time of these modes at elevated temperatures, smoothening away the distinct vibrational lines present at 77 K. Such vibrational energy redistribution and dephasing reflect the anharmonic nature of these vibrations. The vibrational mode assignments in the 1D system are based on the finite differences calculations of the inorganic framework by excluding the organic component since the inclusion of a large amount of ODA ligands in one unit cell introduces a significant challenge in convergence and increases mode complexity. We thus discuss the lattice vibrations considering the closest correspondence with experimental results. The vibrational modes at 23, 51, and 136 cm$^{-1}$ are attributed to pure rocking, twisting, and scissoring of I–Sn–I, respectively. The other modes represent mixed I–Sn–I motions: twisting with rocking (34 cm$^{-1}$), twisting with asymmetric stretching (77 cm$^{-1}$), I-Sn-I wagging with asymmetric stretching (104 cm$^{-1}$), see Fig. 4a, Supplementary Fig. S29, and Table S7.

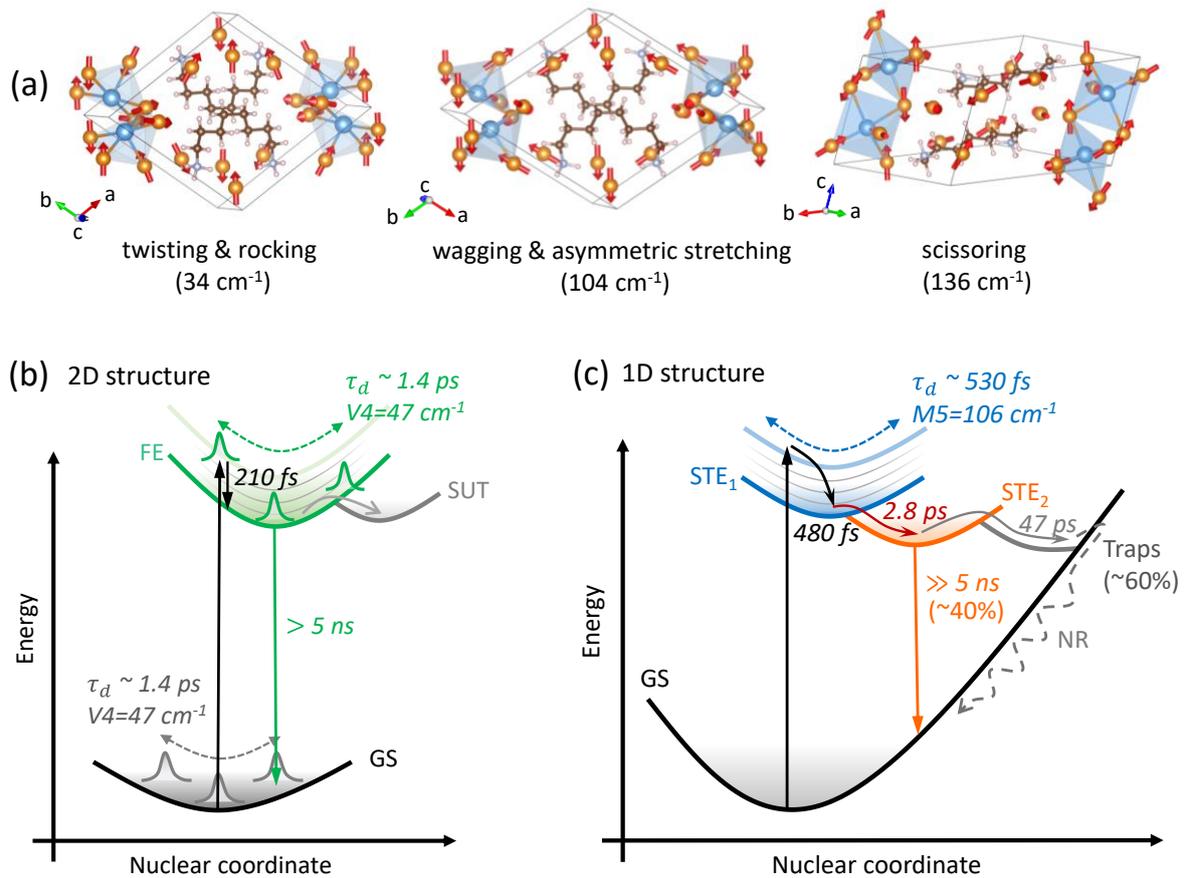

**Fig. 4 Summary of the vibrations, wavepacket dynamics and electronic processes.** (a) Selection of calculated vibrational modes and the corresponding atomic motions in ODASn$_2$I$_6$. Schematic illustrations of vibrational and electronic dynamics: (b) 2D ODASnI$_4$ and (c) 1D ODASn$_2$I$_6$ thin films. The black upward arrow indicates ultrashort pulse excitation initiating coherent wavepackets on the electronic states via ISRS (oscillation on the ground state) or DECP (oscillation on the excited state). The damping time of each dominant mode is denoted by $\tau_d$ with the mode frequencies labeled M for the 1D system and V for the 2D system. The dominant mode M5 of the 1D system is assigned to the 104 cm$^{-1}$ mode in the calculations. Dashed double arrows depict oscillatory wavepacket propagation. In the 2D structure, the high-energy excitation populates the hot FE state that undergoes carrier cooling (black downward arrow). Light grey and green arrows represent carrier trapping to form the surface trap state (SUT) and the second order recombination, respectively. In the 1D structure, the dark downwards arrow corresponds to the initial rapid dynamics which consist of carrier cooling and self-trapped exciton formation combined with wavepacket dynamics. The red arrow shows the electronic relaxation between the two STE states. Grey and yellow arrows show that ~60% of excitations decay nonradiatively, while the remaining ~40% contribute to PL emission.

**Discussion**

The coherent phonons can be generated either displacively in the excited state or impulsively in the ground electronic state. In both cases, to be visible in TA as oscillations, the potential energy surfaces (PES) of the corresponding modes need to be displaced. Initiation of the excited state wavepacket is thereby sometimes called displacive excitation of coherent phonons (DECP). Since the phonons start at maximum amplitude, they show cosine-like oscillations.[36, 46, 44] The ground state wavepacket is explained by the impulsive stimulated Raman scattering (ISRS).[22, 47, 41, 42] In this case the wavepacket starts with negligible initial amplitude and thereby shows sine-like oscillations. The initial phase of coherent oscillations indicates which excitation pathway dominates in the excitation of the wavepacket.[44] As shown in Supplementary Fig. S30a, the oscillatory signal at 595 nm in the 2D system presents an initial phase of $\pi$, which falls in the Franck-Condon region evolving on the excited state PES. Along with the propagation of the excited state wavepacket, the phonon oscillation at 550 nm shifted by $\pi$ out of phase relative to that at 595 nm, indicating that the wavepacket has passed through the equilibrium position on the PES of the FE state and is now oscillating on the opposite side. Additionally, the V4 mode at 595 nm (GSB) and 550 nm (ESA) presents the phase of $1.8\pi$ and $0.64\pi$ which are in between the sine- and cosine-like oscillation most likely originating from their combination (Supplementary Table S3). In our experiments, the 2D sample was excited above the bandgap, then the vibrational wavepackets could form on both the ground and excited state. It means that the oscillations in the 2D system generally originate from both ISRS and DECP. The schematic vibrational dynamics are shown in Fig. 4b. The fitted dominant mode frequencies (V4) at 595 nm and 550 nm have a difference of 2 cm$^{-1}$ falling in the range of frequency uncertainty (3 cm$^{-1}$), therefore we present that the vibrational wavepackets propagating along the PES of the ground state and FE state are the same. In TA measurements of the 1D system, we only observe the ESA signals. The oscillatory signals at 500 and 851 nm are almost perfectly in phase (Supplementary Fig. S30b). Taken together, this suggests that the vibrational wavepackets reside on the excited state induced via DECP. In the oscillatory signal analyses, we notice that the M5 mode (106 cm$^{-1}$) is dominant with the highest amplitude. The Fourier transform of the bandgap fluctuation autocorrelation function is dominated by a mode of very similar frequency (104 cm$^{-1}$), which gives the largest contribution to the reorganization energy. Therefore, we conclude that the corresponding vibrational wavepackets propagate along the STE formation potential energy surface, and the M5 mode is related to the lattice distortion, driving the system to be strongly self-trapping, see Fig. 4c.

In summary, we have elucidated that ligand concentration causes the material to have either the 1D or 2D structure with distinct exciton-phonon interactions, leading to the very different electronic and vibrational dynamics in these two materials. Our findings reveal that the STE population dynamics remain largely unaffected by changes in excitation conditions, however, the FE dynamics are significantly influenced by hot carrier cooling and high-order recombination, particularly at higher excitation intensities. The ultrafast pump-probe measurements demonstrate that the coherent vibrational wavepackets on STE state in the 1D system are dominated by a phonon mode at 106 cm$^{-1}$ involved in the wagging and asymmetric stretching vibration in [SnI$_5$]$^-$, while the coherent vibrational wavepackets on FE state and the ground state in the 2D system have significantly lower frequencies below 50 cm$^{-1}$. This distinction underscores that the dimensionality-dependent exciton-phonon interaction acts as an important factor in modulating both vibrational dynamics and electronic transitions, such as exciton self-trapping in low-dimensional organic-inorganic metal halides and inspires a straightforward yet powerful strategy for tailoring their optoelectrical properties.

**Acknowledgments**

This work was supported by the Swedish Energy Agency (Grant 50667-1, 50709-1), the Swedish Research Council VR (2021-05207, 2023-05244), Olle Engkvist Foundation (Grant 235-0422), Knut and Alice Wallenberg Foundation (Dnr KAW 2019.0082) and the China Scholarship Council (No. 202006150002). The computations were enabled by resources provided by the National Academic Infrastructure for Super-Computing in Sweden (NAISS) via the project 2024/5-372. Collaboration with NanoLund is acknowledged. Y. W. acknowledges the Postgraduate Research & Practice Innovation Program of Jiangsu Province (Grant No. KYCX23_0368). J. C. acknowledges the Novo Nordisk Foundation (Grant No. NNF22OC0073582).

**Competing interests**

The authors declare no competing interests.

**Methods**

**Thin film preparation and characterization**

The organic ligands are synthesized according to the previous report.[26] Octylenediammonium diodide (ODA-$I_2$) and $SnI_2$ were dissolved in anhydrous DMF at volume ratios of 2:1 or 4:1 to prepare precursor solutions with a concentration of 1 mol/mL. The nm-scale thin films are spin-coated from the prepared precursor on a clean substrate and annealed at 75°C for 8 minutes. The more details are illustrated in the Supplementary. The film's general morphology was characterized using a Philips XL30FEG scanning electron microscope (SEM) operated at 3 kV. Powder X-ray diffraction (XRD) measurements were performed on a Pananalytical X'Pert Pro diffractometer equipped with a Cu Kα X-ray source (λ = 1.5406 Å). Steady-state absorption spectra were recorded using a UV-vis spectrophotometer (PerkinElmer Lambda 1050). Photoluminescence (PL) spectra and time-correlated photokinetics were obtained with a Horiba Spex 1681 spectrometer, using 375 nm excitation. For temperature-dependent PL, measurements were taken during heating from 77 to 298 K. PL decay dynamics of ODA$Sn_2I_6$ were analyzed with a custom-built time-correlated single-photon counting (TCSPC) setup, featuring Silicon single-photon avalanche diodes (Si SPADs, IDQ100, quantum sensing), a time-tagging device (quTAG), and a 375 nm pulsed laser (EPLED Series, 100 kHz to 2.5 MHz). The system's time resolution was approximately 40 ps. PL decay dynamics of ODA$SnI_4$ were analyzed home-built TCSPC setup with a 405 nm picosecond pulsed laser (laser diodes, LDH-series, PicoQuant) for excitation. The adjustable repetition rate is 2.5 MHz with a pulse duration of 20 ps. The single photon avalanche diode (SPAD, PicoQuant) is used as a detector with a timing resolution as short as 50 ps.

**Femtosecond transient absorption measurements**

Femtosecond to nanosecond transient absorption (fs-ns TA) measurements were carried out using a home-built femtosecond pump-probe setup. The system utilized laser pulses (8 W, 796 nm, 60 fs, 4 kHz) generated by a Solstice amplifier (Spectra Physics) seeded by a Mai Tai SP femtosecond oscillator (Spectra Physics). The seedling laser was split into pump and probe beams. A thin $CaF_2$ plate was used to generate supercontinuum white light for the probe, covering the visible spectrum from 480 to 900 nm. The pump pulse (70 mW, 400 nm, 100 fs) was produced by a collinear optical parametric amplifier (TOPAS-C, Light Conversion) through the frequency doubling of 800 nm. The polarization angle between the pump and probe beams was set to the magic angle (54.7°). For temperature-dependent fs-TA measurements, the sample was mounted in the Oxford cryostat with vacuum conditions to avoid moisture. The excitation fluences for $ODASn_2I_6$ and $ODASnI_4$ are ~ $1.2×10^{16}$ and ~ $2.0×10^{14}$ excitation/$cm^2$/pulse. Global analysis (GLA) was performed using the Glotaran software package (http://glotaran.org), employing singular value decomposition (SVD) for global fitting incorporating multiple components.[48]

**Computational details**

Density functional theory (DFT) calculations were performed using the Projector Augmented Wave (PAW) method to solve the Kohn-Sham equations, as implemented in the Vienna Ab initio Simulation Package (VASP).[49, 50] The generalized gradient approximation (GGA) of Perdew, Burke, and Ernzerhof (PBE) was employed to describe the exchange and correlation terms within the Kohn-Sham Hamiltonian by using the non-spin-polarized approach.[50] Plane waves were expanded with an energy cutoff of 520 eV. The Brillouin zone sampling for the electronic structure calculations was performed using a reciprocal grid of 2×2×3. For structural optimization, a 1×1×2 reciprocal grid was also employed. The force convergence criterion was set to 0.001 eV/Å, while the energy convergence threshold was set to $10^{-5}$ eV. To mitigate the effects of electron self-interaction and to prevent band gap underestimation, the hybrid Heyd–Scuseria–Ernzerhof (HSE06) functional was used.[51] Phonon modes were computed using the finite differences approach and the vibrational modes were analyzed at the Γ point. During this phase, energy convergence was set to $10^{-8}$ eV with a plane-wave cutoff of 400 eV. Ab initio molecular dynamics (AIMD) simulations were performed at 77 K, 200 K, and 298 K for the NVT ensemble. The VACF simulation runs over 5 ps with a time step of 1 fs. The time-correlated energy bandgap simulation runs over 2 ps with a time step of 20 fs.